\documentclass[fleqn,usenatbib]{mnras}

\usepackage{mathptmx}

\usepackage[T1]{fontenc}
\usepackage{ae,aecompl}

\usepackage{graphicx} 
\usepackage{amsmath} 

\usepackage{amssymb} 
\usepackage{amsmath}
\usepackage[utf8]{inputenc}
\usepackage{mathptmx}
\usepackage{breqn}
\usepackage{booktabs}
\usepackage{subfig}
\usepackage{subcaption}
\usepackage{scalerel}
\usepackage{tikz}
\usetikzlibrary{svg.path}

\definecolor{orcidlogocol}{HTML}{A6CE39}
\tikzset{
  orcidlogo/.pic={
    \fill[orcidlogocol] svg{M256,128c0,70.7-57.3,128-128,128C57.3,256,0,198.7,0,128C0,57.3,57.3,0,128,0C198.7,0,256,57.3,256,128z};
    \fill[white] svg{M86.3,186.2H70.9V79.1h15.4v48.4V186.2z}
                 svg{M108.9,79.1h41.6c39.6,0,57,28.3,57,53.6c0,27.5-21.5,53.6-56.8,53.6h-41.8V79.1z M124.3,172.4h24.5c34.9,0,42.9-26.5,42.9-39.7c0-21.5-13.7-39.7-43.7-39.7h-23.7V172.4z}
                 svg{M88.7,56.8c0,5.5-4.5,10.1-10.1,10.1c-5.6,0-10.1-4.6-10.1-10.1c0-5.6,4.5-10.1,10.1-10.1C84.2,46.7,88.7,51.3,88.7,56.8z};
  }
}

\newcommand\orcidicon[1]{\href{https://orcid.org/#1}{\mbox{\scalerel*{
\begin{tikzpicture}[xscale=1,yscale=-1, transform shape]
\pic{orcidlogo};
\end{tikzpicture}
}{|}}}}

\usepackage{hyperref}
\hypersetup{urlcolor=blue, colorlinks=true}  

\title[Spin Evolution in CC Scl]{Evolution of Spin in the Intermediate Polar CC Sculptoris}

\author[John A. Paice et al.]{John A. Paice$^{\orcidicon{0000-0003-1149-1741}}$,$^{1}$\thanks{E-mail: \href{mailto:johnapaice@gmail.com}{johnapaice@gmail.com}}
    S. Scaringi$^{\orcidicon{0000-0001-5387-7189}}$,$^{1}$
    N. Castro Segura$^{\orcidicon{0000-0002-5870-0443}}$,$^{2}$
    A. Sahu$^{\orcidicon{0009-0007-6825-3230}}$,$^{2}$
    K. I\l{}kiewicz$^{\orcidicon{0000-0002-4005-5095}}$$^{1,3}$,\newauthor
    Deanne L. Coppejans$^{\orcidicon{0000-0001-5126-6237}}$,$^{2}$
    D. De Martino$^{\orcidicon{0000-0002-5069-4202}}$,$^{4}$
    C. Knigge$^{\orcidicon{0000-0002-1116-2553}}$,$^{5}$
    and M. Veresvarska$^{\orcidicon{0000-0002-0146-3096}}$$^{1}$
	\\
	$^{1}$Centre for Extragalactic Astronomy, Department of Physics, Durham University, South Road, Durham DH1 3LE, UK\\
    $^{2}$Department of Physics, University of Warwick, Gibbet Hill Road, Coventry CV4 7AL, UK\\
    $^{3}$Astronomical Observatory, University of Warsaw, Al. Ujazdowskie 4, 00-478 Warszawa, Poland\\
    $^{4}$INAF-Osservatorio Astronomico di Capodimonte, Salita Moiariello 16, I-80131 Naples, Italy\\
    $^{5}$Department of Physics and Astronomy. University of Southampton, Southampton SO17 1BJ, UK
}

\date{Submitted to MNRAS in original form 2024 AAA XX, Received 2024 AAA XX, Accepted 2024 AAA XX}

\pubyear{2024}

\begin{document}
\label{firstpage}
\pagerange{\pageref{firstpage}--\pageref{lastpage}}
\maketitle

\begin{abstract}
    We report on spin variations in the intermediate polar and cataclysmic variable CC Scl, as seen by the Transiting Exoplanet Survey Satellite (TESS). By studying both the spin period and its harmonic, we find that the spin has varied since it was first observed in 2011. We find the latest spin value for the source to be 389.473(6)\,s, equivalent to 0.00450779(7) days, 0.02\,s shorter than the first value measured. A linear fit to these and intermediate data give a rate of change of spin (\.P) $\sim$ -4.26(2.66)$\times$10$^{-11}$ and a characteristic timescale $\tau$ $\sim$ 2.90$\times$10$^{5}$ years, in line with other known intermediate polars with varying spin. The spin profile of this source also matches theoretical spin profiles of high-inclination intermediate polars, and furthermore, appears to have changed in shape over a period of three years. Such `spin-up' in an intermediate polar is considered to be from mass accretion onto the white dwarf (the primary), and we note the presence of dwarf nova eruptions in this source as being a possible catalyst of the variations.

	
\end{abstract}

	



\section{Introduction} \label{sec:Intro}

Cataclysmic Variables (CVs) are compact binary systems where a white dwarf accretes from a late-type companion star via Roche Lobe overflow. This accretion is strongly affected by the white dwarf's magnetic field: When the white dwarf's magnetic field is low, it accretes from a disc; however, for white dwarfs with strong magnetic fields (>10$^6$\,G), matter is swept up along the magnetic field lines and channelled onto the poles.

In the latter, if the accretion occurs directly from the stream from the companion, then the system will not host a disc, and the white dwarf's spin will synchronise with the orbit; these systems are known as polars. However, in weaker magnetic regimes, the magnetic field will instead merely truncate the disc, sweeping up matter at the `magnetospheric radius'; such systems are known as intermediate polars.

The interaction between the magnetic field and the disc can transfer angular momentum both to the white dwarf (during accretion) and from it (e.g. the propeller effect, \citealt{WynnKingHorne_PropellorAEAqr_1997}), respectively increasing or decreasing the white dwarf's spin. Spin variations have been seen in several intermediate polar CVs, such as DQ Her, AO Psc, FO Aqr, V1223 Sgr, and BG CMi (see \citealt{Patterson_IPSpinHistory_2020} for details).

One particular intermediate polar, CC Sculptoris (also known as CC Scl), was first discovered as an X-ray source (RX J2315.5-3049, \citealt{Schwope_CCSclDiscovery_2000}) and later also found to undergo dwarf nova eruptions \citep{Ishioka_CCSclFirstOptical_2001}. These eruptions occur when the accretion disc becomes ionised, similar to many low-mass X-ray binary systems (see \citealt{Lasota_DiscInstabilityModel_2001, Done_EverythingAccretion_2007} for reviews). 


\citet{Woudt_CCSclSpin_2012} confirmed CC Scl as an intermediate polar with a spin period of 389.49\,s (0.00450801 days). \citet{Kato_CCScl_2015} and \citet{Szkody_CCScl+HST_2017} investigated the spin since, with the latter being the first to detect it during quiescence. Both \citet{Woudt_CCSclSpin_2012} and \citet{Szkody_CCScl+HST_2017} noted that the harmonic of the spin period (at twice the frequency) is sometimes stronger than the fundamental; the authors of the former suggested (and the latter supported) that this is due to the second pole being visible, as CC Scl is a high-inclination system \citep{Kato_CCScl_2015}. All the spin determinations of this source so far are shown in Table \ref{tab:spins}.


We now present new data on the variability of this source, obtained from data taken by the Transiting Exoplanet Survey Satellite (TESS), which has observed CC Scl on two separate occasions. We investigate the frequency of the spin and harmonic, compare it with previous data, and quantify the changes that may have happened over the past decade.


    


\begin{table}
	\centering
    \caption{Historical spin determinations for this source.}
    \begin{tabular}{l l l l}
    \toprule 
    \textbf{MJD} & \textbf{Spin (s)} & \textbf{Spin (days)} & \textbf{Reference} \\
    \midrule

    55717 & 389.492(2) & 0.00450801(6) & 1 \\
    55874 & 389.457(20) & 0.0045076(2) & 2 \\
    56472 & 387.371 & 0.0044835 & 3$^{\dagger}$ \\
    56850 & 389.483(80) & 0.0045079(9) & 2 \\

    \bottomrule
    \end{tabular}
    \vspace{0.2cm}
    
    \footnotesize{$^1$\citet{Woudt_CCSclSpin_2012}}; 
    \footnotesize{$^2$\citet{Kato_CCScl_2015}}; 
    \footnotesize{$^3$\citet{Szkody_CCScl+HST_2017}}
     
    \footnotesize{$^\dagger$See Appendix}\\
	\label{tab:spins}
\end{table}

\section{Method} \label{sec:Method}

\begin{figure*}
\centering
\includegraphics[width=\textwidth]{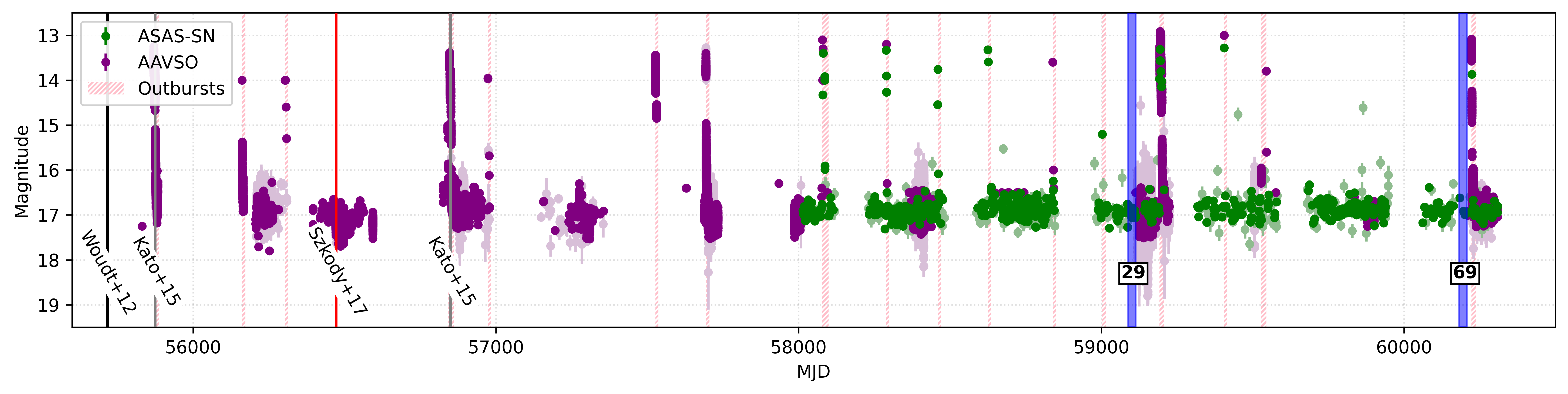}
\caption{Long-term light curves of CC Scl from ASASSN (green) and AAVSO (purple), with outbursts (arbitrarily defined as magnitude < 16) in red. Faded points have error > 0.1 mag, and are not considered for outbursts. Previous observations in literature \citep{Woudt_CCSclSpin_2012, Kato_CCScl_2015, Szkody_CCScl+HST_2017} and the two TESS sectors investigated in this paper (29 and 69) are also plotted.}
\label{fig:ASASSN_LC}
\end{figure*}

\begin{figure}
\centering
\includegraphics[width=\columnwidth]{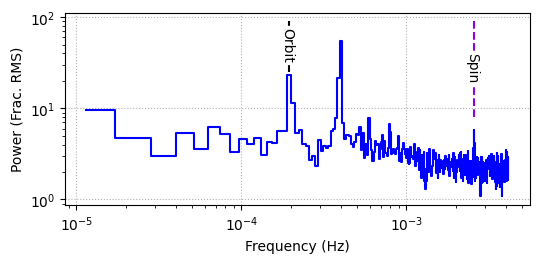}
\caption{Power spectrum of TESS sector 69. The orbit and its harmonic are strongest, with the spin at higher frequencies.}
\label{fig:PSD}
\end{figure}

Data were taken by TESS, the Transiting Exoplanet Survey Satellite \citep{Ricker_TESS_2015}, which observes in the optical/near-infrared bands, between $\sim$600--1000\,nm. The two datasets were from Sector 29 (2020 August 26 -- 2020 September 21) and Sector 69 (2023 August 25 -- 2023 September 20), and the long-cadence (`LC') data (with a cadence of 120\,s) were obtained and analysed using the \texttt{Lightkurve} python package\footnote{\href{https://docs.lightkurve.org/index.html}{https://docs.lightkurve.org/index.html}}. We used the SAP (Simple Aperture Photometry) flux as our data, and any rows with a quality flag > 0 were removed.

Figure \ref{fig:ASASSN_LC} shows the ASAS-SN (All-Sky Automated Survey for Supernova, \citealt{Shappee_ASASSN_2014, Kochanek_ASASSN_2017}) and AAVSO (American Association of Variable Star Observers) light curves of the source; dwarf nova outbursts can be seen to happen regularly, with a recurrence time of around 175 days. Previous investigations into CC Scl's spin, as well as the relevant TESS sectors for this paper, are also plotted. A power spectrum of sector 69 of the TESS data, averaged over 20 segments ($\sim$1 day in length), can be seen in Figure \ref{fig:PSD}. 

To investigate the spin of CC Scl, we used a Lomb-Scargle periodogram \citep{lomb_least-squares_1976, scargle_studies_1982}. We only investigated the region around the fundamental spin frequency (221.80--221.87 cycles/day), outside of which there were no significant signals. The Lomb-Scargle package we used was from \texttt{astropy}\footnote{\href{https://docs.astropy.org/en/stable/index.html}{https://docs.astropy.org/en/stable/index.html}}, and with it we used twenty `samples per peak' as the oversampling.


In order to find the uncertainties of the peak of the Lomb-Scargle, we also employed bootstrapping, wherein N points were drawn from the original light curves with replacement (where N is the length of the original light curve), and the above analyses were carried out on the resultant data. To find the peak of the signal, the highest bin in this range was taken, along with two bins on either side; a polynomial was fit to them, and the location of the maximum of that polynomial was recorded as the maximum frequency. This was done 50,000 times. Afterwards, a histogram of the frequencies was recorded, and a Gaussian was fit; the mean of this Gaussian was taken to be the mean value of the signal, and the standard deviation of the population of bootstrapped points was taken to be the error.

We also considered the harmonic of the spin at twice the frequency, also noted in \citet{Woudt_CCSclSpin_2012} and in several subsequent papers. For this region, the frequency interval 443.64--443.72 cycles/day was used, outside of which there were no significantly high peaks. Lomb-Scargle analysis of this harmonic required going beyond the Nyquist frequency, and as such a Nyquist factor of 2 was used; this is possible because any signal beyond the Nyquist frequency is reflected back into the sub-Nyquist range, and shows a peak at the relevant frequency (which we see).


\section{Results} \label{sec:Results}

\begin{figure*}
\begin{tabular}{cc}
\subfloat{\includegraphics[width = 0.5\textwidth]{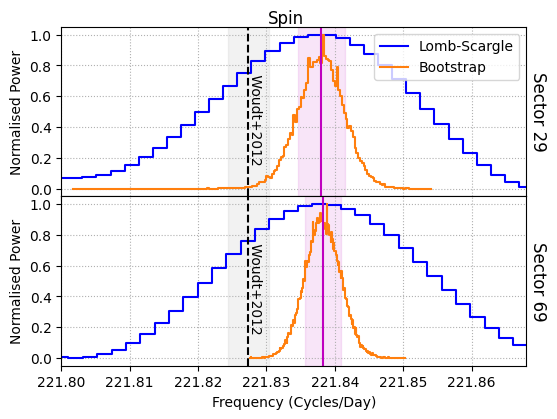}} &
\subfloat{\includegraphics[width = 0.5\textwidth]{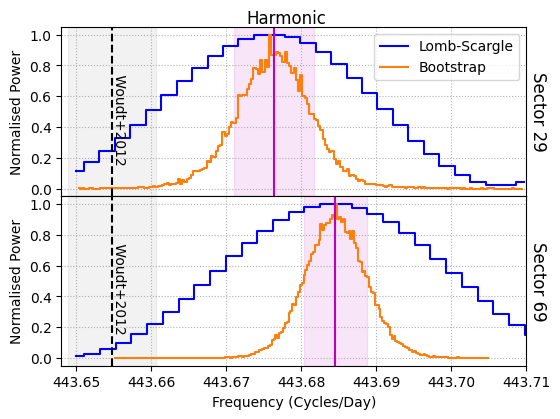}}
\end{tabular}
\caption{Lomb-Scargle periodograms (blue) of CC Scl about the spin period, for sectors 29 (top) and 69 (bottom). The peaks of bootstrapped Lomb-Scargle (left) and PDM (right) analyses are shown in orange, with their mean and standard deviation shown in purple. Each analysis has been normalised to unity. Also plotted in black is the spin period and error reported in \citet{Woudt_CCSclSpin_2012}.}\label{fig:spin_period}
\end{figure*}

Lomb-Scargles for the spin and harmonic are plotted in Figure \ref{fig:spin_period}, alongside the bootstrapped analysis, its mean, and its standard deviation (i.e. one-sigma uncertainty), and a comparison is made with the spin period given by \citet{Woudt_CCSclSpin_2012}. Values for the spin and the harmonic, as well as the harmonic translated to the spin, are shown in Table \ref{tab:our_spins}.

Both measurements of the fundamental are consistent. For the harmonic, there is a small increase in the frequency, but this is not outside of errors. However, each case shows a strong deviation from the original spin period reported in \citet{Woudt_CCSclSpin_2012}.

\begin{table}
	\centering
    \caption{Spin and Harmonic determinations from this work, with one-sigma uncertainties. `Equivalent Spin' is our Harmonic measurements multiplied by a factor of 2.}
    \begin{tabular}{l l l l l}
    \toprule 
    \textbf{MJD} &
    \textbf{Period} &
    \vline &
    \textbf{Equivalent Spin} &
    \\
     & 
    (s) &
    \vline &
    (s) &
    (days) \\
    \midrule

    59100 & 389.4733(60) & \vline & 389.4733(60) & 0.004507792(69) \\
    60194 & 389.4730(46) & \vline & 389.4730(46) & 0.004507789(53) \\
    \hline
    59100 & 194.7365(23) & \vline & 389.4731(47) & 0.004507790(54) \\
    60194 & 194.7329(18) & \vline & 389.4659(36) & 0.004507707(42) \\

    \bottomrule
    \end{tabular}
	\label{tab:our_spins}
\end{table}

\section{Discussion} \label{sec:Discuss}
\begin{figure}
\centering
\includegraphics[width = \columnwidth]{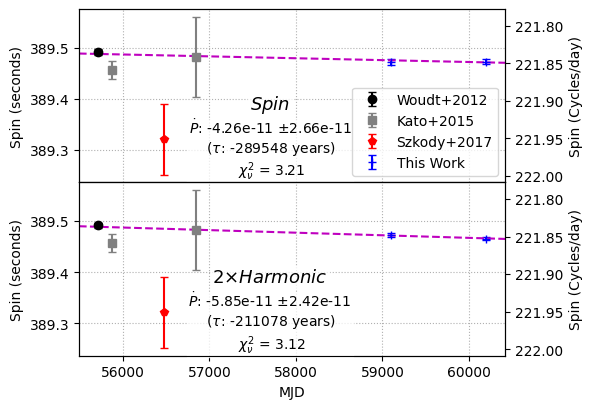}
\caption{Plot of the spin determination over time, including a linear fit to calculate the rate of change of spin (\.{P}) and characteristic timescale ($\tau$). The top plot fits to our spin measurements, and the bottom plot fits to twice our harmonic measurements. We also fit to the values from \citet{Woudt_CCSclSpin_2012}, \citet{Kato_CCScl_2015}, and  \citet{Szkody_CCScl+HST_2017}. Note that the spin period from \citet{Szkody_CCScl+HST_2017} is calculated from their reported harmonic; see Appendix.}\label{fig:spin_change_linear}
\end{figure}

We plot our results with those from \citet{Woudt_CCSclSpin_2012}, \citet{Kato_CCScl_2015}, and \citet{Szkody_CCScl+HST_2017} in Figure \ref{fig:spin_change_linear}, and in each case carry out a $\chi^2$ fit. \citet{Szkody_CCScl+HST_2017}'s point used the harmonic; see Appendix for a discussion of those data.

It is unclear what the overall trend is for this source. The points from \citet{Kato_CCScl_2015} and \citet{Szkody_CCScl+HST_2017} imply a sudden, strong increase in spin of 0.2\,s over at minimum one year (\.P$\sim$10$^6$). Intermediate polars do vary their spin around equilibrium (see, e.g., \citealt{Warner_SpinUpDown_1990, Patterson_DQHer_1994, Norton_SpinPeriods_2004}), but not on timescales like this (\citealt{Patterson_IPSpinHistory_2020}, see also Table \ref{tab:lit_spins}); it should also be noted that these points have large uncertainties. We do note, though, that the spin of CC Scl is not inconsistent with it being in stable equilibrium; P$_{spin}$ $\sim$ 0.078 P$_{orb}$ \citep{Woudt_CCSclSpin_2012}, a value consistent with the theory of disc-like intermediate polars \citep{NortonParker_AccFlowsOfMagCVs_2008} and the general population \citep{WynnKing_IPs_1995}.

However, what is clear is that there has been some shift between the original measurement in 2011 by \citet{Woudt_CCSclSpin_2012} and the two most recent TESS sectors in 2020 and 2023. A simple linear fit to the data gives a spin-up rate between -4.26(2.66)$\times$10$^{-11}$ for the spin and -5.85(2.42)$\times$10$^{-11}$ for the harmonics, to one sigma, with reduced $\chi^{2}$ of 3.21 and 3.12 respectively. This equates to a characteristic timescale ($\tau$, equal to the spin over \.P) of $\sim$2.90$\times$10$^{5}$ and $\sim$2.11$\times$10$^{5}$ years respectively. This is similar to other intermediate polar systems with variable spins; see Table \ref{tab:lit_spins}.


\begin{table}
	\centering
    \caption{Spins from five other Intermediate Polars that show (or have at one point shown) linear spin-changes, including their rate of change of spin (\.P) and characteristic timescale ($\tau$), compared with our values.}
    \begin{tabular}{l l l l l}
    \toprule 
    \textbf{Name} &
    \textbf{Spin} (s) &
    \textbf{\.P} &
    \textbf{$\tau$} (years) &
    \textbf{Ref.}
    \\
    
    \midrule
    PQ Gem & 833.42 & 9.08(5)$\times$10$^{-11}$ & 2.91$\times$10$^{5}$ & 1 \\
    V418 Gem & 240.34 & -3.0(2)$\times$10$^{-12}$ & -2.52$\times$10$^{6}$ & 2 \\
    GK Per & 351.33 & 1.2(2)$\times$10$^{-11}$ & 8.57$\times$10$^{5}$ & 3 \\
    AO Psc & 858.62 & -5.93$\times$10$^{-11}$ & -5.05$\times$10$^{5}$ & 4 \\
    V1223 Sgr & 794.41 & 2.35$\times$10$^{-11}$ & 1.07$\times$10$^{6}$ & 4 \\
    \textbf{CC Scl} & \textbf{389.49} & \textbf{-4.26$\times$10$^{-11}$} & \textbf{-2.90$\times$10$^{5}$} & \textbf{-} \\

    \bottomrule
    \end{tabular}
    \vspace{0.2cm}
    \\ \footnotesize{$^1$\citet{Evans_PQGem_2006}}; \footnotesize{$^2$\citet{Patterson_V418Gem_2011}};\\ \footnotesize{$^3$\citet{Zemko_GKPer_2017}}; \footnotesize{$^4$\citet{Patterson_IPSpinHistory_2020}}\\
	\label{tab:lit_spins}
\end{table}

\subsection{Spin Profile}

\begin{figure}
\centering
\includegraphics[width = \columnwidth]{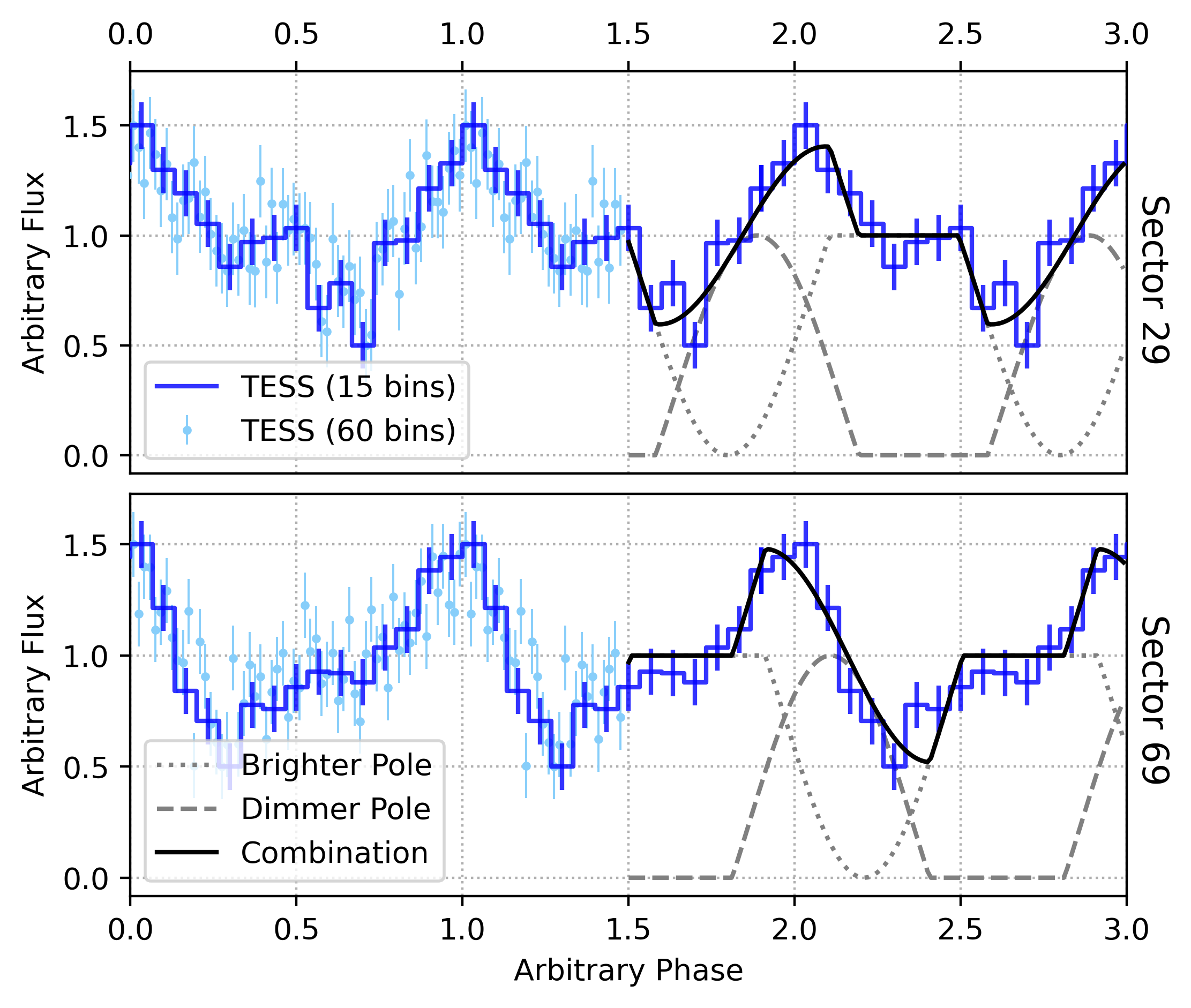}
\caption{Light curves of sector 29 (top) and 69 (bottom) folded on the mean spin value for sector 29 shown in Table \ref{tab:our_spins}. Three periods are shown for clarity. Two resolutions are shown; one averaged over 60 bins (light blue, left) and one over 15 bins (dark blue). A model similar to that suggested by \citet{Hellier_XYAri_1997} is also fitted and is plotted in black; two asymmetric components (the brighter and dimmer accretion poles, dotted and dashed lines respectively) combine to create a complex profile. Both model and fit are representative only, meant to show the changing shape of the spin pulse profile between sectors and its relation to the expected light curve of two asymmetric accretion poles.}\label{fig:folded_light_curves}
\end{figure}

To investigate the source further, we found the spin phase profile for both sectors. We did this by folding the light curves on the spin period for sector 29 (since the spins are within 1 sigma of each other); see Figure \ref{fig:folded_light_curves}. T$_0$ was chosen independently for each sector, as our ephemeris was not good enough to phase across sectors. 

The folded light curves bear a resemblance to XY Ari, a similarly high-inclination IP, as shown in \citet{Hellier_XYAri_1997}. 
In that source, a combination of the upper accretion zone (the one tilted towards us) and the lower accretion zone (the one tilted away from us), when combined with some asymmetry between the two, give rise to a three-step shape: a plateau when just the upper accretion region is visible; then a peak while it's turning away and the lower accretion region is coming into view; and finally a dip as the lower region turns away and the upper region is still coming into view. We created a representative, phenomenological model matching this idea: two sinusoids, with some cut-off value (one at the top, to represent the upper pole, and one at the bottom, to represent the lower), are offset slightly in phase and then combined. The resultant is fitted to our CC Scl data; this is shown in the right half of Fig. \ref{fig:folded_light_curves}; note that this is a representative model and fit only, merely to show that our data is consistent with this concept. This supports previous determinations by \citet{Woudt_CCSclSpin_2012} and \citet{Szkody_CCScl+HST_2017} that both poles are visible, and their combination produces the harmonic. 

The spin pulse profile of CC Scl changes between sectors; the peak occurs after the dip in sector 29, and before the dip in sector 69. This is reminiscent of `pole-switching', which occurs when accretion favours one pole and then the other; although such effects are primarily seen in asynchronous polars (e.g. CD Ind, \citealt{Littlefield_SpinOrbitCDInd_2019}, and BY Cam, where \citealt{Mason_BYCamMagneticValve_2022} ascribe it to a magnetic valve at L1 modulating the accretion flow).

\section{Conclusion} \label{sec:Conclusion}

CC Scl is an intermediate polar that has shown a significant change in its spin over the course of several years. While intervening measurements have large uncertainties, the first spin measurement from 2011 \citep{Woudt_CCSclSpin_2012} is significantly different from the most recent determinations made in 2020 and 2023; fitting to all measurements implies that the source is spinning up with a \.P of -4.26(2.66)$\times$10$^{-11}$ and a characteristic timescale of $\sim$2.90$\times$10$^{5}$ years. 



Further observations of this source in the coming years would be highly desirable, and TESS is already planning to observe this source in a future sector. Other bands, such as X-rays, could also test the two-pole model and monitor the evolution of the spin period.


\section*{Acknowledgements}
	
JAP acknowledges support from STFC consolidated grant ST/X001075/1. AS acknowledges the
Warwick Astrophysics PhD prize scholarship made possible thanks to a generous philanthropic donation. KI acknowledges support from Polish National Science Center grant Sonatina 2021/40/C/ST9/00186. MV acknowledges the support of
the Science and Technology Facilities Council (STFC) studentship
ST/W507428/1.
This paper includes data collected with the TESS mission, obtained from the MAST data archive at the Space Telescope Science Institute (STScI). Funding for the TESS mission is provided by the NASA Explorer Program. STScI is operated by the Association of Universities for Research in Astronomy, Inc., under NASA contract NAS 5–26555. We acknowledge with thanks the ASAS-SN team, as well as variable star observations from the \textit{AAVSO International Database} contributed by observers worldwide and used in this research. This research is based on observations made with the NASA/ESA Hubble Space Telescope obtained from the Space Telescope Science Institute, which is operated by the Association of Universities for Research in Astronomy, Inc., under NASA contract NAS 5–26555. These observations are associated with program 12870 (PI: Gaensicke).

\section*{Data Availability Statement}

Data from the TESS and HST missions are publicly available from the Mikulski Archive for Space Telescopes (MAST). Data from the ASAS-SN are publicly available from Sky Patrol\footnote{\href{https://asas-sn.osu.edu}{https://asas-sn.osu.edu}}. Data from AAVSO are publicly available from their Data Access page\footnote{\href{https://www.aavso.org/data-access}{https://www.aavso.org/data-access}}.

\bibliography{ref}
\bibliographystyle{mnras}
	

	
	
\appendix

\section{Hubble Space Telescope data from Szkody+2017} \label{sec:Discuss_Szkody}

\citet{Szkody_CCScl+HST_2017} presented ultraviolet data on CC Scl as observed by Hubble Space Telescope (HST) using the Cosmic Origins Spectrograph \citep{COS}, on 2013 June 29 at 03:39:03~UT with an effective exposure time of 1.3~h covering two HST orbits, as part of the GO programe 12870 \citep{2012hst..prop12870G}. In this work, \citet{Szkody_CCScl+HST_2017} found the spin period peak at around 387\,s, and a stronger harmonic peak at 194.657\,s, but did not give an error on either.

We downloaded the HST data from the Mikulski Archive for Space Telescopes (MAST). Light curves were extracted from the time-tag event list following the method described in \citet{Castro-Segura2022Natur.603...52C}, and geocoronal spectral regions were removed from the event list to avoid artefacts in the light curves. To test the influence of the emission lines in the resulting periodograms, we also created light curves from spectral regions with no clear emission line contribution. The continuum regions were selected by visual inspection from the average spectrum: the regions lie between $\lambda\lambda = (1352.0,1377.6), (1425.1,1516), (1573,1622), (1665,1839)$ and $(1870,2000)$~\AA.

\begin{figure}
\includegraphics[width = 0.5\textwidth]{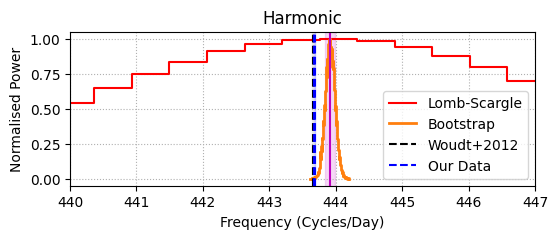}
\caption{Analysis of HST data first presented in \citet{Szkody_CCScl+HST_2017}; Lomb-Scargle (red) and subsequent bootstrap (orange) of the harmonic period, with mean and standard deviation presented in purple. Our and Woudt's data are shown.}\label{fig:szkody_spin_period}
\end{figure}

In both cases, we found the spin period to be far less significant than the harmonic, which explains the deviation in reported spin seen in Table \ref{tab:spins}. Instead, we focused on the harmonic; we applied a Lomb-Scargle periodogram and bootstrapped the result between 435--450 cycles, with everything else the same as described in Section \ref{sec:Method}. We find the harmonic to be 194.63(3)\,s for the full light curve and 194.66(4)\,s for the continuum, which is equivalent to 389.26(6) and 389.32(7) respectively (note that each are consistent within uncertainties). The latter case is shown in Figure \ref{fig:szkody_spin_period}.



However, the breadth of the Lomb-Scargle peak should be noted. In Figure \ref{fig:szkody_spin_period}, the locations of both \citet{Woudt_CCSclSpin_2012} and our spins are very close to the peak of the Lomb-Scargle of the HST data, even if the bootstrapping does not align with them. Figure 5 in \citet{Szkody_CCScl+HST_2017} shows that the window size for the discrete Fourier transform is much broader than the spin differences seen in this source. Alongside the much shorter data train for this data compared with TESS, investigating the HST light curve shows a lot of stochastic variability in the source, which is an additional complication for determining the spin.

\bsp 
\label{lastpage}
\end{document}